\documentclass[12pt]{article}
\usepackage{amssymb}
\usepackage{amsmath}
\usepackage{amsthm}
\usepackage[space]{cite}
\usepackage{xcolor}

\usepackage{graphicx} 

\numberwithin{equation}{section}

\newcommand{\id}{\mathbb{I}}

\begin{document}
\begin{titlepage}

\begin{center}
\vspace*{\skip5}

{\LARGE $su(2)$ symmetry of XX spin chains}\\
\vspace{1.cm}

{\large Nicolas Cramp\'e$^{a}$, Rafael I. Nepomechie$^b$, 
Luc Vinet$^{c,d,e}$ \footnote{email:luc.vinet@umontreal.ca (Corresponding author)}\\ \vspace{0.5cm} and Nabi Zare Harofteh$^b$}

\vspace{1.cm}

$^a$CNRS – Universit\'e de Montr\'eal CRM - CNRS, France\\\vspace{0.5cm}
$^b$Department of Physics, PO Box 248046\\
University of Miami, Coral Gables, FL 33124 USA\\ \vspace{0.5cm}
$^c$Centre de Recherches Math\'ematiques, Universit\'e de Montr\'eal\\
P.O. Box 6128, Centre-ville Station\\
Montr\'eal (Qu\'ebec), H3C 3J7, Canada\\\vspace{0.5cm}
$^d$D\'epartement de Physique, Universit\'e de Montr\'eal\\ 
Montr\'eal (Qu\'ebec), H3C 3J7, Canada\\\vspace{0.5cm}
$^e$IVADO, 6666 Rue Saint-Urbain\\
Montr\'eal (Qu\'ebec), H2S 3H1, Canada

\end{center}

\vspace{.5in}

\begin{abstract}
We show that, after suitably adjusting a uniform transverse magnetic field, the generic inhomogeneous open XX spin chain has a two-fold degeneracy, and an exact $su(2)$ symmetry whose ``inhomogeneous'' nonlocal generators depend on coefficients that can be explicitly computed for models associated with discrete orthogonal polynomials.
\end{abstract}

\end{titlepage}

\setcounter{footnote}{0}

\section{Introduction}\label{sec:intro}

The XX spin chain has been widely studied, see e.g. \cite{Lieb:1961fr, Barouch:1970ryz, Crampe:2019upj, Finkel:2020lgf, Bernard:2024eoj} and references therein.
We consider here the inhomogeneous open XX spin-chain Hamiltonian 
defined on a lattice with $N+1$ sites\footnote{Our notation closely follows that of \cite{Crampe:2019upj}.}
\begin{equation}
 \mathcal{H} = -\frac{1}{2}\sum_{n=0}^{N-1}J_{n} (\sigma^x_n\sigma^x_{n+1}+ \sigma^y_n\sigma^y_{n+1})-\frac{1}{2} \sum_{n=0}^{N}B_n \sigma^z_{n}\,,
 \label{HXX}
\end{equation}
where $J_n \ne 0$ and $B_n$ are real parameters.
This Hamiltonian has the evident $U(1)$ symmetry
\begin{equation}
	\left[ \mathcal{H} \,, S^{z}  \right] =0  \,, \qquad
S^{z}  = \frac{1}{2}\sum_{n=0}^{N} \sigma^{z}_{n}\,.
\label{Sz}
\end{equation}	
In this note, we observe that, for arbitrary values of the parameters $J_n$ and $B_n$ (up to a possible shift of the latter parameter by an $n$-independent constant, which corresponds to turning on a uniform magnetic field in the $z$ direction), this Hamiltonian possesses the additional exact symmetry
\begin{equation}
	\left[ \mathcal{H} \,, \mathbb{S}^{x}  \right] = \left[ 
	\mathcal{H} \,, \mathbb{S}^{y}  \right] 
    = \left[ \mathcal{H} \,, \mathbb{S}^{z}  \right]  =0   \,,
    \label{symmetry}
\end{equation}
where  $\mathbb{S}^{x}$ and $\mathbb{S}^{y}$ are corresponding ``inhomogeneous'' nonlocal generators
of the form
\begin{equation}
	\mathbb{S}^{x} = \frac{1}{2}\sum_{n=0}^{N} a_{n}\, 
    \sigma^z_0\dots \sigma_{n-1}^z \sigma_n^x \,,	\qquad
	\mathbb{S}^{y} = \frac{1}{2}\sum_{n=0}^{N} a_{n}\, 
    \sigma^z_0\dots \sigma_{n-1}^z \sigma_n^y \,,
	\label{SxSy}
\end{equation}
whose coefficients $a_{n}$ (which depend on the model parameters $J_n$ and $B_n$) follow from a general expression given below,
see \eqref{coefficients}.
Explicit analytic expressions for these coefficients can be obtained for models associated with discrete orthogonal polynomials. Moreover,
\begin{equation}
    \mathbb{S}^z := -i\left[ \mathbb{S}^{x} \,,  \mathbb{S}^{y}  
	\right] \,.
    \label{SSz}
\end{equation}
The algebra generated by $\mathbb{S}^{x}$, $\mathbb{S}^{y}$ and $\mathbb{S}^{z}$ is $su(2)$, with its Casimir element given by
\begin{equation}
    \vec{\mathbb{S}}^{\, 2} =(\mathbb{S}^{x})^2+(\mathbb{S}^{y})^2+(\mathbb{S}^{z})^2=\frac{3}{4}\mathbb{I}\,,
\end{equation}
which follows from $\sum_{n=0}^N(a_n)^2=1$, proven in the following. This result shows that this representation of the $su(2)$ algebra is a direct sum of two-dimensional irreducible representations. This symmetry implies that the spectrum of the Hamiltonian is
two-fold degenerate.

We shall see that the $su(2)$ symmetry and two-fold degeneracy of the spectrum can be understood from the fact that the corresponding fermionic system has a zero-energy mode. 

Generators of $su(2)$ that depend on model parameters and are spatially dependent appeared recently in \cite{Znidaric:2024mwj}. However, the system considered there is homogeneous, and the symmetry is only up to boundary terms. The generators \eqref{SxSy} resemble inhomogeneous versions of so-called semilocal charges considered in e.g. \cite{Fagotti:2021efb, Fagotti:2022npd}.

We present the general construction of the symmetry generators and their algebra in Section \ref{sec:gen}. Two examples are worked out in Section \ref{sec:examples}. We conclude with a brief discussion in Section \ref{sec:discussion}.

\section{Fermionic zero modes and symmetry generators}\label{sec:gen}

In order to prove the symmetries \eqref{symmetry} and determine the coefficients $a_{n}$ in the ``inhomogeneous'' generators \eqref{SxSy}, we use the Jordan--Wigner transformation \cite{Jordan:1928wi}
\begin{equation}
    c_n^\dagger =\sigma^z_0\dots \sigma_{n-1}^z 
\sigma_n^+ \,, \qquad
    c_n =\sigma^z_0\dots \sigma_{n-1}^z \sigma_n^- \,,
    \qquad \sigma^\pm = \frac{1}{2}\left(\sigma^x \pm i \sigma^y\right) \,,
\label{JW}
\end{equation}
where $\{ c_{m}^{\dagger} \,, c_{n} \} = \delta_{m,n}$
and $\{ c_{m}\,, c_{n} \} =0$,
to bring the Hamiltonian \eqref{HXX} to that of an inhomogeneous free-fermion model with open boundary conditions
\begin{equation}
\mathcal{H}=\sum_{n=0}^{N-1}J_{n} (  c_n^\dagger c_{n+1} + c_{n+1}^{\dagger} c_{n})- \sum_{n=0}^{N}B_n c_{n}^{\dagger} c_{n} \,,
\label{eq:Hff}
\end{equation}
up to an additive constant.
The Hamiltonian \eqref{eq:Hff} can be rewritten in quadratic form
\begin{equation}
\mathcal{H}=\sum_{m,n=0}^N c^\dagger_m  H_{mn}c_n \,,
\label{eq:Hff2}
\end{equation}
where $H$ is the $(N+1)\times (N+1)$ tridiagonal matrix 
\begin{equation}
 H=\begin{pmatrix}
             -B_0 & J_0 & \\
             J_0 & -B_1 & J_1 \\
             & J_1 & -B_2 & J_2 \\
             && \ddots & \ddots & \ddots \\
            &&&J_{N-2} & -B_{N-1} & J_{N-1}\\
             &&&& J_{N-1} & -B_N
            \end{pmatrix}\,.
\label{Hmat}
\end{equation}

We denote by $|\omega_k\rangle$
the orthonormal eigenvectors of $H$ with eigenvalues $\omega_k$
\begin{equation}
 H |\omega_k\rangle = \omega_k |\omega_k\rangle\,, \qquad
 k = 0, 1, \ldots, N \,.
\end{equation}
A key point is that one of the eigenvalues of $H$, say $\omega_K$, can always be set to 0
\begin{equation}
    \omega_{K} = 0 \,.
    \label{zeromode}
\end{equation}
Indeed, this can evidently be arranged by performing the constant shift $H \mapsto  H - \omega_{K} \id$, where $\id$ is the identity matrix, which results in a shift of all the eigenvalues 
$\omega_k \mapsto \omega_k - \omega_{K}$.
In view of \eqref{Hmat}, this means
shifting $B_n \mapsto B_n +  \omega_{K}$, which corresponds in \eqref{HXX} to turning on the uniform magnetic field $-\omega_{K} \hat{z}$.
The eigenvectors can be expressed as
\begin{equation}
 |\omega_k\rangle =\sum_{n=0}^N \phi_n(\omega_k) |n\rangle\,,
 \label{phi}
\end{equation}
where $\{|0\rangle,|1\rangle,\dots ,|N\rangle \} $ is the 
canonical orthonormal basis of $\mathbb{C}^{N+1}$. The eigenfunctions 
$\phi_n(\omega_k)$ are real, since the matrix $H$ is 
real and its eigenvalues are non-degenerate (see e.g. Lemma 3.1 in 
\cite{Terwilliger:2004}). These eigenfunctions therefore satisfy the orthonormality conditions
\begin{equation}
\sum_{n=0}^N \phi_n(\omega_k) \phi_n(\omega_p)=\delta_{kp}\,.
\label{orthophi}
\end{equation}

The free-fermion Hamiltonian 
$\mathcal{H}$ \eqref{eq:Hff} becomes diagonal 
\begin{equation}
\mathcal{H}=\sum_{k=0}^{N} \omega_{k} 
\tilde{c}^{\dagger}_{k} \tilde{c}_{k} 
\label{Hdiag}
\end{equation}
when expressed in terms of the annihilation operators $\tilde{c}_{k}$ defined by
\begin{equation}
\tilde{c}_{k} = \sum_{n=0}^{N} \phi_{n}(\omega_{k})\, 
c_{n} \,, \qquad
c_{n} = \sum_{k=0}^{N} \phi_{n}(\omega_{k})\, 
\tilde{c}_{k} \,,
\label{ctilde}
\end{equation}
and the creation operators  
$\tilde{c}^{\dagger}_{k}$ given by 
Hermitian conjugation of \eqref{ctilde}, hence 
$\{ \tilde{c}^{\dagger}_{k} \,, \tilde{c}_{p} \} = \delta_{k,p}$ 
and $\{ \tilde{c}_{k} \,, \tilde{c}_{p} \} = 0$.
In view of \eqref{Hdiag},
the eigenvectors of $\mathcal{H}$ are given by
\begin{equation}
	|\Psi\rangle\!\rangle = \tilde{c}_{k_{1}}^\dagger \ldots \tilde{c}_{k_{r}}^\dagger |0\rangle\!\rangle \,,
	\label{grounstate}
\end{equation}
with corresponding energy eigenvalues $E=\sum_{i=1}^{r} \omega_{k_{i}}$,
where $k_{1}, \ldots, k_{r} \in \{0, \ldots, N\}$ are pairwise distinct, and
the vacuum state $|0\rangle\!\rangle$ is annihilated by all the annihilation 
operators
\begin{equation}
	\tilde{c}_{k}|0\rangle\!\rangle = 0\,, \qquad k = 0\,, \ldots\,, 
	N \,.
\label{vacuum}
\end{equation}

We now observe that, due to the zero mode \eqref{zeromode}, the Hamiltonian \eqref{Hdiag} commutes with the corresponding zero-mode operators
\begin{equation}
    \left[ \mathcal{H} \,, \tilde{c}^\dagger_{K} \right] =  \left[ \mathcal{H} \,, \tilde{c}_{K} \right] =  0 \,.
\end{equation}
It follows that the XX spin-chain Hamiltonian \eqref{HXX} indeed has the claimed symmetry \eqref{symmetry}, where the symmetry generators $\mathbb{S}^x$ and $\mathbb{S}^y$ are given by linear combinations of the fermionic zero-mode operators
\begin{equation}
    \mathbb{S}^x = \frac{1}{2}\left(\tilde{c}^\dagger_{K} + \tilde{c}_{K}\right) \,, 
    \qquad \mathbb{S}^y = -\frac{i}{2}\left(\tilde{c}^\dagger_{K} - \tilde{c}_{K} \right)\,;
    \label{SxSyzeromode}
\end{equation}
the $a_n$ coefficients in \eqref{SxSy} are given, in view of \eqref{JW} and \eqref{ctilde}, by
\begin{equation}
a_n = \phi_n(\omega_K) = \phi_n(0)  \,.
    \label{coefficients}
\end{equation}

The generators $\mathbb{S}^{x}$,  $\mathbb{S}^{y}$ and $\mathbb{S}^{z}$ 
(the latter defined in \eqref{SSz}) close to an $su(2)$ algebra
\begin{equation}
	\left[ \mathbb{S}^{x} \,, \mathbb{S}^{z} \right]  = -i \mathbb{S}^{y}  \,, \qquad
	\left[ \mathbb{S}^{y} \,, \mathbb{S}^{z} \right]  = i \mathbb{S}^{x}  \,.
\end{equation}
If we extend this algebra to include $S^{z}$ \eqref{Sz}, then we have
\begin{align}
	\left[ \mathbb{S}^{x} \,, S^{z} \right]  &=  - i \mathbb{S}^{y}  \,, \label{more1} \\
	\left[ \mathbb{S}^{y} \,, S^{z} \right]  &=   i  \mathbb{S}^{x}  \,, \label{more2} \\
	\left[ \mathbb{S}^{z}  \,, S^{z} \right]  &=  0 \,.
    \label{more3}
\end{align}
We further note the identities
\begin{equation}
	 \left( \mathbb{S}^{x} \right)^{2} = \left( \mathbb{S}^{y} \right)^{2} =   \left( \mathbb{S}^{z} \right)^{2} =  \frac{1}{4}\id \,,    
    \label{identities}
\end{equation}
which follow from \eqref{SxSyzeromode}.

The symmetry \eqref{symmetry} implies that the spectrum of the Hamiltonian
has a two-fold degeneracy. Indeed, if $|\Psi\rangle\!\rangle$ is an eigenvector of $\mathcal{H}$, then  $\mathbb{S}^x \, |\Psi\rangle\!\rangle$ is also an eigenvector of $\mathcal{H}$ with the same eigenvalue. Moreover, $|\Psi\rangle\!\rangle$  and 
$\mathbb{S}^x \, |\Psi\rangle\!\rangle$ are linearly independent\footnote{Recalling \eqref{Sz}, let $|\Psi\rangle\!\rangle$ be a simultaneous eigenvector of 
$\mathcal{H}$ and $S^z$; and let us assume that 
$\mathbb{S}^x \, |\Psi\rangle\!\rangle$  and $|\Psi\rangle\!\rangle$ are {\it not} linearly independent, so that
$$\mathbb{S}^x \, |\Psi\rangle\!\rangle = \alpha\,  |\Psi\rangle\!\rangle$$
for some constant $\alpha$. Applying  \eqref{more1} to $|\Psi\rangle\!\rangle = |\Psi\rangle\!\rangle$, we obtain 
$\mathbb{S}^y \, |\Psi\rangle\!\rangle = 0$; and it follows from \eqref{identities} that $|\Psi\rangle\!\rangle = 0$, which is a contradiction.},
and $(\mathbb{S}^x)^2$ 
is proportional to the identity operator.

\section{Examples}\label{sec:examples}

We present here two examples with the symmetry 
\eqref{symmetry}-\eqref{SSz},
where the coefficients $a_n$ \eqref{coefficients} can be computed explicitly. Both examples correspond to models associated with discrete orthogonal polynomials \cite{Crampe:2019upj}.

\subsection{Homogeneous chain}

For the homogeneous chain, the Hamiltonian is given by \eqref{HXX} with
\begin{equation}	
J_{0}= \ldots = J_{N-1} = -\frac{1}{2} \,, \qquad B_{0}= \ldots = B_{N-1} 
= -\cos\left(\frac{\pi (K+1)}{N+2}\right) \,,
\label{JBCheb}
\end{equation}
for which the eigenfunctions in \eqref{phi} are given by \cite{Crampe:2019upj}
\begin{equation}	
\phi_n(\omega_k) = \sqrt{\frac{2}{N+2}} \sin \left[\frac{\pi 
(k+1)(n+1)}{N+2}\right]\,, 
\label{phiCheb}
\end{equation}
where
\begin{equation}
\omega_k = \cos\left(\frac{\pi (K+1)}{N+2}\right)
-\cos\left(\frac{\pi (k+1)}{N+2} \right) \,, \qquad k = 0, 1, \ldots, N 
\,.
\end{equation}
Note that $B_n$ in \eqref{JBCheb} have been adjusted to ensure the zero mode $\omega_K=0$.
The coefficients $a_n$ \eqref{coefficients} are therefore given by
\begin{equation}
    a_n = \sqrt{\frac{2}{N+2}} \sin \left[\frac{\pi 
(K+1)(n+1)}{N+2}\right] \,.
\end{equation}
One can check, by brute-force diagonalization for $N=1, 2 \ldots$ and $K=0, 1, \ldots, N$, that all the eigenvalues of the Hamiltonian \eqref{HXX} with parameters \eqref{JBCheb} are (at least) two-fold degenerate.

\subsection{Krawtchouk chain}

For the Krawtchouk chain, the Hamiltonian is given by \eqref{HXX}
with \cite{Crampe:2019upj}
\begin{equation} 
	J_{n} =  \sqrt{(N-n)(n+1)p(1-p)}\,, \qquad
	B_{n} = -\left[N p + n(1 - 2 p)\right]\,, \qquad 0<p<1 \,.
    \label{JBKrawt}
\end{equation}
The eigenfunctions in \eqref{phi} are given by
\begin{equation}
\phi_n(\omega_k) =  (-1)^{n} 
\sqrt{\left(\frac{p}{1-p}\right)^{n+k} (1-p)^N{N\choose n} {N\choose k}}\,
R_{n}(-k) \,,
\end{equation}
where $R_{n}(-k)$ are Krawtchouk polynomials that can be expressed in terms of the hypergeometric function
\begin{equation}
	\setlength\arraycolsep{1pt}
	R_{n}(-k) = {}_{2} F_{1}\left(\begin{matrix} -n, & & -k
	\\&-N  &\end{matrix}\ 
	;\frac{1}{p}\right)\,, \qquad n = 0, 1, \ldots, N\,,
	\label{Krawtchouk}
\end{equation}	
and
\begin{equation}
\omega_k = k \,, \qquad k= 0, 1, \ldots, N \,.
\end{equation}
Note that here $\omega_0 = 0$, corresponding to the choice $K=0$.
The coefficients $a_n$ \eqref{coefficients} are therefore given by
\begin{equation}
	a_{n} = (-1)^{n} p^{\frac{n}{2}} (1-p)^{\frac{1}{2}(N-n)}
	\sqrt{{N\choose n}} \,.
\end{equation}	
One can explicitly check that all the eigenvalues of the Hamiltonian \eqref{HXX} with parameters \eqref{JBKrawt} and $N=1, 2 \ldots$ are (at least) two-fold degenerate.

\section{Discussion}\label{sec:discussion}

We have seen that, after suitably adjusting a uniform transverse magnetic field,
the generic inhomogeneous open XX spin chain has a two-fold degeneracy, and an enhanced $su(2)$ symmetry \eqref{symmetry}. The ``inhomogeneous'' symmetry generators \eqref{SxSy} depend on coefficients $a_n$ \eqref{coefficients} that can be computed explicitly for models associated with discrete orthogonal polynomials. The expressions for the generators \eqref{SxSy}, with their non-local ``tails'' of $\sigma^z$ operators, are reminiscent of coproducts of quantum-group generators \cite{Chari:1994pz}. 
The exact two-fold degeneracy of all the levels of the open XX chain due to the $su(2)$ symmetry presented here is reminiscent of the approximate pairing of all the levels in the ordered phase of the one-dimensional open Ising chain due to the presence of a strong zero mode \cite{Kitaev:2000nmw, Fendley:2012vv, Kemp:2017ifs}. The edge spins of the latter model are known to have long coherence times \cite{Kemp:2017ifs}. For the former model, perhaps some (presumably nonlocal) observable could be found, which has suitable overlap with the $su(2)$ charges and is therefore conserved for long times. The symmetry uncovered here may provide further insight into 
quench dynamics \cite{Fagotti:2021efb, Fagotti:2022npd} and 
protected phases \cite{deBuruaga:2018mvc} in inhomogeneous spin chains.
It would be very interesting if this type of symmetry could be extended to XXZ-type chains.
It would also be interesting to find higher-rank algebras with such ``inhomogeneous'' generators, and corresponding physical models with such symmetries.

\section*{Acknowledgements}

We thank Eric Vernier for helpful correspondence.
NC is supported in part by the international research project AAPT of
the CNRS. RN is supported in part by the National Science Foundation under grant PHY 2310594, and by a Cooper fellowship.  LV is funded in part by a Discovery Grant from the Natural Sciences and Engineering Research
Council (NSERC) of Canada. 

\section*{Conflict of interest}
On behalf of all authors, the corresponding author states that there is no conflict of interest.

\section*{Data availability}
This manuscript has no associated data.


\providecommand{\href}[2]{#2}\begingroup\raggedright\endgroup

\end{document}